\documentclass[a4paper]{article}
\usepackage{graphicx}
\usepackage{multirow}

\newcommand\pubnumber{}
\newcommand\pubdate{}

\def\pisa{INFN, Sezione di Pisa\\
Largo B. Pontecorvo 3, 56127 Pisa, ITALY}

\def\Title#1{\begin{center} {\Large #1 } \end{center}}
\def\Author#1{\begin{center}{ \sc #1} \end{center}}
\def\Address#1{\begin{center}{ \it #1} \end{center}}

\newcommand\pubblock{\rightline{\begin{tabular}{l} \pubnumber\\
         \pubdate  \end{tabular}}}
\newenvironment{Abstract}{\begin{quotation}  }{\end{quotation}}
\newenvironment{Presented}{\begin{quotation} \begin{center} 
             PRESENTED AT\end{center}\bigskip 
      \begin{center}\begin{large}}{\end{large}\end{center} \end{quotation}}
\def\Acknowledgements{\bigskip  \bigskip \begin{center} \begin{large}
             \bf ACKNOWLEDGEMENTS \end{large}\end{center}}

\def\beq{\begin{equation}}
\def\eeq#1{\label{#1}\end{equation}}
\def\eeqn{\end{equation}}

\def\beqa{\begin{eqnarray}}
\def\eeqa#1{\label{#1}\end{eqnarray}}
\def\eeqan{\end{eqnarray}}

\let\bar=\overbar

\def\Dslash{\not{\hbox{\kern-4pt $D$}}}
\def\dslash{\not{\hbox{\kern-2pt $\del$}}}

\def\msb{{\bar{\ssstyle M \kern -1pt S}}}

\def\bxsg {\ensuremath{B\to X_s\gamma}}
\def\bxdg {\ensuremath{B\to X_d\gamma}}
\def\bxsll {\ensuremath{B\to X_s\ell^+\ell^-}}
\def\mxs {\ensuremath{m_{X_s}}}
\def\acp {\ensuremath{A_{CP}}}
\def\dacp {\ensuremath{\Delta A_{CP}}}
\def\BB {\ensuremath{B{\bar B}}}
\def\epem {\ensuremath{e^+e^-}}
\def\mpmm {\ensuremath{\mu^+\mu^-}}
\def\qsq {\ensuremath{q^2}}
\def\afb {\ensuremath{A_\mathrm{FB}}}
\newcommand\result[5] {\ensuremath{ #1 ^{+ #2 + #4}_{- #3 - #5} } }

\begin{document}
\begin{titlepage}
\pubblock

\vfill
\Title{Inclusive $B\to X_s\gamma$ and $B\to X_s\ell^+\ell^-$ at the B factories}
\vfill
\Author{ John Walsh}
\Address{\pisa}
\vfill
\begin{Abstract}
I report here recent measurements of observables from the inclusive decays $B\to X_s\gamma$ and $B\to
X_s\ell^+\ell^-$. Included are measurements of the branching fractions and CP asymmetries for both channels, as well as the
forward-backward lepton asymmetry in inclusive $B\to X_s\ell^+\ell^-$ decays, which is the first measurement of this
quantity.
\end{Abstract}
\vfill
\begin{Presented}
FPCP 2014 -- Flavor Physics \& CP Violation\\
Marseille, France,  May 26--30, 2014
\end{Presented}
\vfill
\end{titlepage}
\def\thefootnote{\fnsymbol{footnote}}
\setcounter{footnote}{0}

\section{Introduction}
\label{sec:intro}

\begin{figure}[htb]
\centering
\includegraphics[height=1.5in]{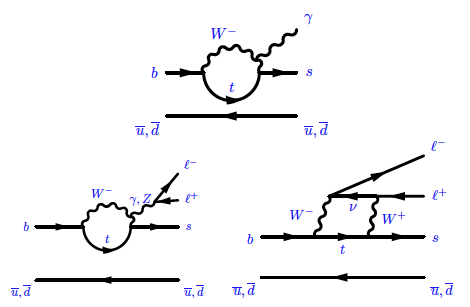}
\caption{Lowest order SM diagrams for \bxsg\ and \bxsll\ decays.}
\label{fig:feynman-diagrams}
\end{figure}

Radiative and electroweak penguin decays, in particular the decays $B\to X_s\gamma$ and $B\to X_s\ell^+\ell^-$, have
proven to be powerful probes of New Physics (NP) in the flavour sector. These flavour-changing neutral current
decays are prohibited at tree level in the Standard Model (SM). This makes them sensitive to NP effects, which can
contribute at the same level as the SM, namely at the one-loop level, as can be seen in Fig.~\ref{fig:feynman-diagrams}.
A general review of radiative and electroweak penguin physics can be found in section 17.9 of
reference~\cite{ref:pbfb}. One usually distinguishes between {\em exclusive} and {\em inclusive} measurements, where in
the former case, the measurement is performed on a particular final state, for example $B^0 \to K^{*0} \gamma$. Recent
results on exclusive measurements were presented at this conference by Patrick Owen and Akimasa
Ishikawa~\cite{ref:exclusive-fpcp2014}. Inclusive analyses attempt to include all final states for a given parton level
process. This has theoretical advantages, since the calculation of inclusive radiative and electroweak
penguin decays is much more precise than the corresponding calculations on exclusive decay modes. In the latter,
hadronic effects tend to cause theoretical uncertainties to grow significantly.

From an experimental point of view, truly inclusive measurements are significantly more challenging: since the $B$ decay
is not fully reconstructed, there are fewer kinematic constraints available in the event selection. Typically, a
fully-inclusive measurement will try to tag one $B$ meson in the event and then look for an inclusive signature of the
signal from the other $B$. An example would be requiring a high-$p_T$ lepton to tag a semi-leptonic $B$ decay and then
require a high-energy photon in the same event, as a signal of the \bxsg\ process.  In such fully inclusive analyses the
backgrounds generally tend to be higher than for exclusive measurements, leading to higher uncertainties.

This difficulty is somewhat alleviated with the {\em sum-of-exclusives} (SOE) technique, whereby a large number
(typically tens) of exclusive final states are reconstructed to capture as much as the full rate as possible. Usually
50--70\% of the total rate is selected and the missing part must be estimated using simulation. This generally leads to
a larger systematic uncertainty than one obtains with the fully inclusive techniques.

In these proceedings, I will report on a measurement of the CP asymmetry in inclusive \bxsg\ decays, using a
fully inclusive method, as well as measurements of the branching fraction and CP asymmetry using the sum-of-exclusives
technique. I will also report measurements of the branching fraction, CP asymmetry and forward-backward (FB) lepton asymmetry
in \bxsll\ decays. The FB lepton asymmetry measurement is the first ever made of this quantity for the inclusive decay. 

All measurements reported were performed either at Belle~\cite{ref:Belle} or Babar~\cite{ref:Babar}, the two B factory
experiments. Each of these detectors operated at an $e^+e^-$ collider operating at a center-of-mass energy of 10.58 GeV,
equal to the mass of the $\Upsilon$(4S) resonance. 

\section{Measurements on inclusive \bxsg}

\subsection{Branching fraction using sum-of-exclusives}

Measurements of the branching fraction of the inclusive \bxsg\ process have been very useful in putting significant
constraints on parameters of models of NP~\cite{ref:bsg-constraints}. A significant portion of the credit for this
success must be attributed to the theorists who have made a precise calculation of the SM branching fraction at
NNLO~\cite{ref:misiak}. The result reads:
\begin{equation}
{\cal B}(\bxsg)|_{E_\gamma>1.6\ \mathrm{GeV}} = (3.15 \pm 0.23) \times 10^{-4}
\end{equation}

Belle reports a preliminary measurement of the inclusive BF using the sum-of-exclusive technique based on their full
$\Upsilon$(4S) dataset of 710 fb$^{-1}$. They fully reconstruct 38 exclusive final states. The hadronic system (labeled
$X_s$) consists of 1 or 3 kaons (at most one $K_S$), up to one $\eta$ and up to four pions (with a maximum of two
$\pi^0$). These states comprise about 70\% of the total rate. The large continuum background is reduced using a
multivariate classifier using the neural network technique. The 12 inputs to this classifier are primarily event-shape
variables, {\em i.e.,} quantities that can separate the more jetty continuum events from the nearly isotropic $B\bar B$
events.

\begin{figure}[htb]
\centering
\includegraphics[height=2.5in]{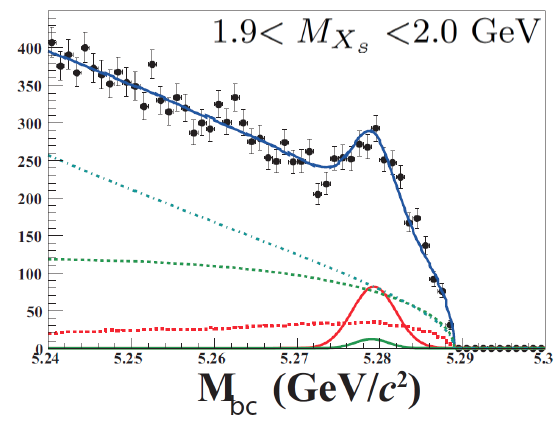}
\caption{Example fit to $M_\mathrm{bc}$ for Belle's \bxsg\ BF measurement. The fit is performed for events with $1.9 <
  M_{Xs} 2.0$ GeV. The signal contribution is the solid red curve. The background contributions (from top to bottom)
  are: continuum, non-peaking \BB, cross feed and peaking \BB, respectively.}
\label{fig:mbc-fit}
\end{figure}

\begin{figure}[htb]
\centering
\includegraphics[height=2.5in]{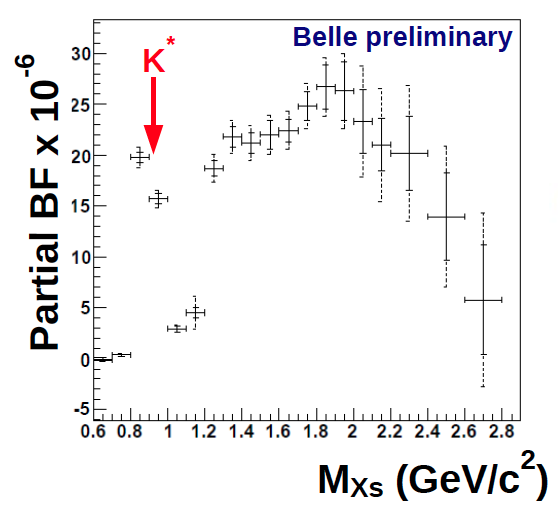}
\caption{Resulting $M_{Xs}$ spectrum for \bxsg. The narrow peaking structure at low \mxs\ is due to the $K*$(892)
  resonance.} 
\label{fig:mxs-spectrum}
\end{figure}

The signal is determined by fitting the beam-constrained $B$ mass, $M_{bc} \equiv \sqrt{E_{\mathrm{beam}}^2 -
    \left|\vec{p}_B\right|^2}$
 in bins of the hadronic system mass \mxs\ from 0.6 to 2.8 GeV. Note that \mxs\ is directly
related to the photon energy in the rest from of the decaying $B$ meson:
\[ 
E_\gamma = \frac{m_B^2 - m_{X_s}^2}{2m_B},
\]
so the quoted mass range corresponds to a photon energy range of 1.9--2.6 GeV. Figure~\ref{fig:mbc-fit} shows an example
fit of $M_{bc}$, for the hadronic mass bin $1.9 < \mxs < 2.0$ GeV. The resulting \mxs\ spectrum, obtained after
performing all the $M_{bc}$ fits, is shown in Figure~\ref{fig:mxs-spectrum}. The narrow peak at around 0.9 GeV
corresponds to the $K^*\gamma$ contribution. The obtained partial branching fraction is:
\begin{equation}
{\cal B}(\bxsg)|_{0.6 < \mxs < 2.8\ \mathrm{GeV}} = (3.51 \pm 0.17_{\mathrm{stat}} \pm 0.33_{\mathrm{syst}}) \times 10^{-4}
\end{equation}
For comparison to the theoretical calculation, this partial rate is extrapolated to $E_\gamma > 1.6$ GeV (equivalent to
$\mxs < 3.31$ GeV), leading to the result:
\begin{equation}
{\cal B}(\bxsg)|_{E_\gamma > 1.6 \mathrm{GeV}} = (3.74 \pm 0.18_{\mathrm{stat}} \pm 0.35_{\mathrm{syst}}) \times 10^{-4}
\end{equation}
The largest contributions to the systematic error are the uncertainty in the fragmentation model and the description of
the $M_{bc}$ probability distribution function (PDF). The result is compatible with the current world average value as
calculated by the Heavy Flavor Averaging Group~\cite{ref:hfag-bsg}:
\begin{equation}
{\cal B}(\bxsg)|_{E_\gamma > 1.6 \mathrm{GeV}}^\mathrm{HFAG} = (3.43 \pm 0.21_{\mathrm{exp}} \pm 0.07_{\mathrm{extrap}}) 
\times 10^{-4}
\end{equation}

\subsection{CP asymmetry using sum-of-exclusives}
Babar has performed a measurement of the CP asymmetry in inclusive \bxsg, using their full dataset of 429
fb$^{-1}$~\cite{ref:babar-acp}. The CP asymmetry is defined as:
\begin{equation}
A_{CP} = \frac{\Gamma({\bar B}\to X_s\gamma) - \Gamma({B}\to X_{\bar s}\gamma)}
 {\Gamma({\bar B}\to X_s\gamma) + \Gamma({B}\to X_{\bar s}\gamma)}.
\end{equation}
This quantity is expected to be small in the SM, with a range of $(-0.6,+2.8)$\%~\cite{ref:benzke}. The authors of this
paper suggest measuring a new quantity, \dacp, which is the difference of $A_{CP}$ measured on charged and neutral $B$
mesons: 
\begin{equation}
\dacp = A_{CP}(B^\pm) - A_{CP}(B^0/{\bar B}^0)
\end{equation}
The authors point out that a measurement of this quantity would give information on the chromo-magnetic dipole Wilson
coefficient $C_8$:
\begin{equation}
\dacp = 4\pi^2\alpha_s\frac{{\tilde{\Lambda}}_{78}}{m_b}\mathrm{Im}\left(\frac{C_8}{C_7}\right),
\end{equation}
where $\tilde{\Lambda}_{78}$ is a hadronic parameter, with a calcualted range of $17 < \tilde{\Lambda}_{78} < 190$
MeV. Since, $C_7$ is essentially known from the BF measurements, measuring \dacp\ would give the first experimental
constraints on $C_8$.  In the SM, where the Wilson coefficients are all real, we have $\dacp = 0$.

The Babar measurement starts with 38 fully-reconstructed exclusive channels, 16 of which are self-tagging and hence used
in the \acp\ measurement.  Photons with center-of-mass energy greater than 1.6 GeV are combined with a hadronic system
havng 1 or 3 kaons, up to 3 pions and 1 $\eta$ particle. Neutral pions and $\eta$s are reconstructed in their
$\gamma\gamma$ decay modes. Charged particle identification is performed to distinguish charged pions and kaons. B
candidates are required to have $0.6 < m_{X_s} < 3.2$ GeV and $|\Delta E| < 0.15$ GeV, where $\Delta E = E_B -
E_\mathrm{beam}$ as measured in the center-of-mass.  Two multi-variate classifiers are employed: one to suppress
continuum backgrounds and the other to select the best candidate in events where multiple candidates have been
identified.

\begin{figure}[htb]
\centering
\includegraphics[width=\textwidth]{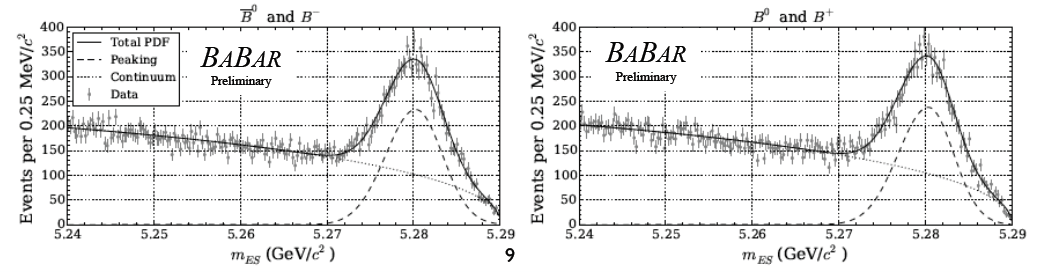}
\caption{Babar \acp\ in \bxsg\ analysis. Fits to $m_\mathrm{ES}$ to candidates containing a $b$ quark (left) or a $\bar
  b$ quark (right).}
\label{fig:babar-acp-mes}
\end{figure}

\begin{figure}[htb]
\centering
\includegraphics[height=2.5in]{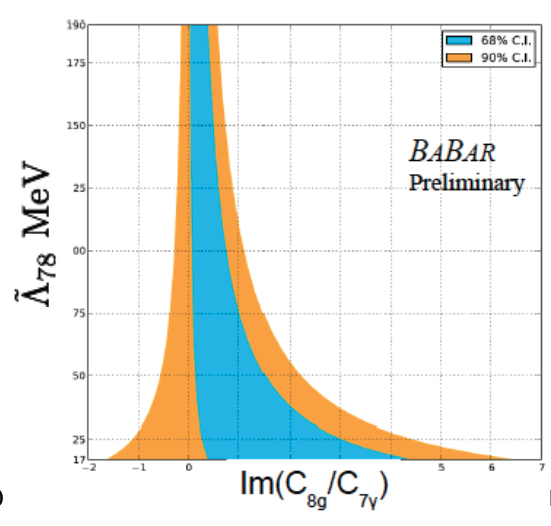}
\caption{Constraint on $\mathrm{Im}(C_8/C_7)$ imposed by this measurement for a given value of $\tilde{\Lambda}_{78}$. The
    current knowledge of $\tilde{\Lambda}_{78}$ corresponds to lowest extremity of the plot.}
\label{fig:acp-constraint}
\end{figure}

Fits to the B candidate mass are performed to extract the yields for $B^+$, $B^-$, $B^0$ and ${\bar B}^0$ decays
(see Fig.~\ref{fig:babar-acp-mes}). 
The resulting raw asymmetries are corrected for inherent detector asymmetry
($A_\mathrm{DET} = (-1.4 \pm 0.7)$\%) and possible background asymmetry ($0.0 \pm 0.9$\%). Combining the charged and
neutral modes together, the full \acp\ is obtained:
\begin{equation}
\acp = (1.7 \pm 1.9_\mathrm{stat} \pm 1.0_\mathrm{syst})\%
\end{equation}
while the simultaneous fit to the charged and neutral samples gives:
\begin{equation}
\dacp = +(5.0 \pm 3.9_\mathrm{stat} \pm 1.5_\mathrm{syst})\%
\end{equation}
This allows us to put the following constraints on the Wilson coefficients:
\[
0.07 \le \mathrm{Im}\frac{C_8}{C_7} \le 4.48, \ 68\%\ \mathrm{CL} \]
\[
-1.64 \le \mathrm{Im}\frac{C_8}{C_7} \le 6.52, \ 90\%\ \mathrm{CL} 
\]
Should the theoretical uncertainty on the hadronic parameter $\tilde{\Lambda}_{78}$ be reduced, the constraint provided
by this measurement will improve substantially as shown in Fig.~\ref{fig:acp-constraint}.

We now turn to a preliminary Belle measurement of the CP asymmetry in fully inclusive \bxsg\ events. As noted above, the
fully inclusive method makes no requirements on the accompanying hadronic system ($X_s$). The basic strategy is to
select a high-energy photon and reduce the substantial continuum background by requiring a high-$p_T$ lepton in the
event, along with some missing energy. This lepton comes from the other B in the event, which has decayed
semileptonically. This lepton tagging method is very effective at reducing the continuum, although it does little to
combat the \BB\ background, where the high-energy photon comes from a B decay that is not \bxsg. A large fraction of the
\BB\ background is removed by vetoing events where the high-energy photon is consistent with the decays
$\pi^0\to\gamma\gamma$ and $\eta\to\gamma\gamma$. The remaining background is then subtracted by using MC
predictions that have been corrected by performing studies on real data. 

One consequence of employing the fully inclusive method is that the event sample will contain the Cabbibo-suppressed
decays \bxdg, as well as \bxsg. For branching fraction measurements, the \bxdg\ component is subtracted from the total
rate. In the case of the CP asymmetry, we cannot do that and we end up measuring the asymmetry for the so-called
``un-tagged'' decay, {\em i.e.}, $B\to X_{s+d}\gamma$. Because of U-spin symmetry, this quantity is almost identically
zero to very high precision~\cite{ref:benzke}, so any significant non-zero measurement of $A_{CP(s+d)}$ would be an
indication of New Physics.

\begin{figure}[htb]
\centering
\includegraphics[height=2.5in]{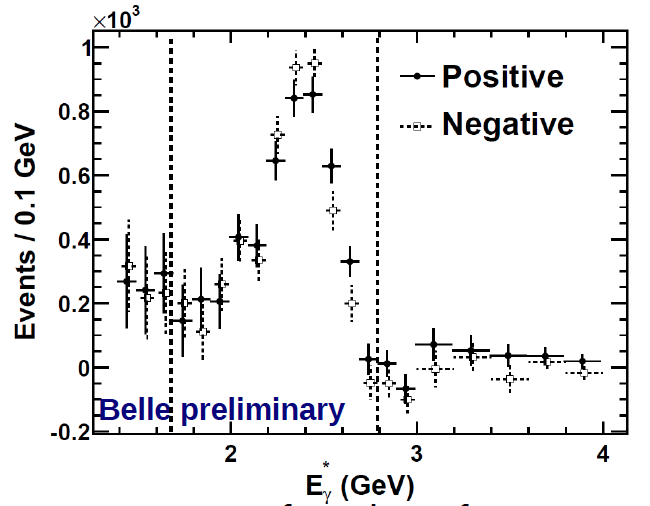}
\caption{The photon energy spectrum for events with a negatively charged (open circles) or positively charged (solid
  circles) lepton. The large peak in the middle is the \bxsg\ signal.}
\label{fig:belle-bsg-acp}
\end{figure}

The sign of the tagging lepton, which contains information on the flavour of the parent B meson, is used to construct
the measured asymmetry:
\begin{equation}
A_{CP}^\mathrm{meas} = \frac{N(\ell^+)-N(\ell^-)}{N(\ell^+)+N(\ell^-)}
\end{equation}
Figure~\ref{fig:belle-bsg-acp} shows the photon energy spectrum for events with a positively or negatively charged
lepton tag. 
The resulting asymmetry \( A_{CP}^\mathrm{meas} = (1.6 \pm 2.9)\times 10^{-2} \) must be corrected for
$B^0{\bar B}^0$ mixing, background and detector effects:
\begin{equation}
A_{CP} = \frac{1}{1-2\omega}A_{CP}^\mathrm{meas} + A_\mathrm{det} + A_\mathrm{bkg}
\end{equation}
This mistag rate $\omega$ arises mostly from mixing, but it is also corrected for cascade decays (where the lepton comes
from a $D$ decay) and for particle misidentification. The mistag rate is found to be: $\omega = 0.1413 \pm 0.0052$. The
detector asymmetry is determined from data using a tag-and-probe method with $B\to X J/\psi(\ell^+\ell^-)$:
$A_\mathrm{det} = (0.1 \pm 0.2)$\%. The asymmetry of the background was determined on events with $E_\gamma^* < 1.7$ GeV
and is found to be $A_\mathrm{bkg} = (-0.1 \pm 0.5)$\%. 

The final result then, with a photon energy cut of 2.1 GeV, reads:
\begin{equation}
A_{CP(s+d)} = (2.2 \pm 4.0_\mathrm{stat} \pm 0.8_\mathrm{syst})\% 
\end{equation}
This result is consistent with zero asymmetry and also with previous measurements of this quantity. 

\section{Measurements on inclusive \bxsll}

The process \bxsll\ is closely related to \bxsg, as can be observed in their Feynman
diagrams (Fig.~\ref{fig:feynman-diagrams}): the photon in the final state is replaced by a pair of leptons. The more complex
final state allows for a wide variety of observables that are sensitive to NP effects, especially observables that
involve an angular analysis of the final state leptons. Furthermore, for most observables, we have theoretical
predictions as a function of the dilepton invariant mass, $q^2 \equiv m_{\ell\ell}^2$, which is a powerful tool for
finding NP and possibly distinguishing among NP models. However, the decay rate for this channel is quite small: about 2
orders of magnitude smaller than the \bxsg\ rate. This makes measurements quite challenging, especially inclusive
measurements. For this reason, few inclusive measurements have been made on \bxsll\ and those have all employed the
experimentally easier sum-of-exclusives method (described above in Sec.~\ref{sec:intro}).

\subsection{Branching fraction and \acp}

Babar has used its full dataset (471 million \BB\ pairs) to measure the branching fraction and CP asymmetry in inclusive
\bxsll\ decays~\cite{ref:babar-sll}. Twenty exclusive final states are reconstructed: 10 different hadronic systems (combinations
of charged or neutral kaons paired with zero, 1 or 2 charged or neutral pions) are
combined with either a pair of muons or electrons. These modes account for about 70\% of the total rate. Kaons, pions,
muons and electrons are selected using particle identification. The kinematics of B decays are exploited, placing
requirements on $\Delta E$ and a multivariate classifier (based on a likelihood ratio LHR) is used to reduce the
background from continuum events.

The event yields are extracted by a 2-dimensional fit to $m_{ES}$ and the LHR. The fits are performed in bins of $q^2$,
which are shown in Table~\ref{tab:q2bins}. 
\begin{table}
\begin{center} 
\label{tab:q2bins}
\caption{The $q^2$ bins used in the Babar analysis of inclusive \bxsll. Note the charmonium veto regions, which are not
  covered by any $q^2$ bin (6.8--10.1 and 12.9--14.2 GeV$^2$ in $q^2$).}
\begin{tabular}{ccc} \hline\hline
$q^2$ bin & $q^2 \equiv m^2_{\ell\ell}$ GeV$^2$ \\ \hline
0 & 1.6--6.0 \\ \hline
1 & 0.2--2.0 \\
2 & 2.0--4.3 \\
3 & 4.3--6.8 \\
4 & 10.1--12.9 \\
5 & 14.2 $-(M_B - M_{K*})^2$ \\ \hline\hline
\end{tabular}
\end{center}
\end{table}
An important aspect of the selection is the veto of charmonium events, {\em i.e.}, the decays $B\to X_s J/\psi$ followed
by $J\psi \to \ell^+\ell^-$. These events have the same final state as the signal events and candidates with $q^2$ in
the range (6.8,10.1) GeV$^2$ and (12.9,14.2) GeV$^2$ are explicitly vetoed. These vetoed events are a very valuable control
sample -- they are used for a wide variety of checks on the simulation of the signal. 

\begin{figure}[htb]
\centering
\includegraphics[width=.8\textwidth]{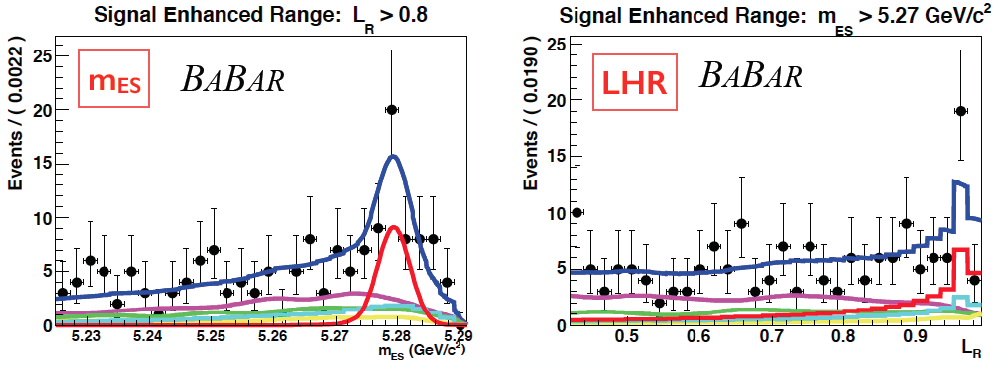}
\caption{Example fit to extract the \bxsll\ rate. The projections in $m_\mathrm{ES}$ (left) and the likelihood ratio
  (right) are shown. The signal is enhanced in each plot by restricting the ``other'' variable to the signal
  region. This fit corresponds to the electron modes in \qsq\ bin 5.}
\label{fig:xee-bin5}
\end{figure}

An example fit for $B\to X_s e^+e^-$ modes in $q^2$ bin 5 is shown in Fig.~\ref{fig:xee-bin5}: on the left is shown the
$m_{ES}$ projection of the fit, while the right plot shows the projection of LHR. For each plot, the signal has been
enhanced by making a loose cut on the other variable. One can see the low statistics which are available in a single
$q^2$ bin. The full set of plots is provided in~\cite{ref:babar-sll} and its supplementary material. 

The derived branching ratios from the event yields have been determined for the \epem\ and \mpmm\ modes separately, as
well as for the combination, in each of the 6 \qsq\ bins. These results are presented in~\cite{ref:babar-sll}, although
here we show the results for which the best theoretical predictions are available~\cite{ref:bsll-theory}, {\em i.e.},
\qsq\ bins 0 and 5.
\begin{table}
\begin{center}
\caption{Branching fraction results in two \qsq\ bins in units of $10^{-6}$. The SM theory calculations are from
  reference~\cite{ref:bsll-theory}. See text for explanation of the quoted uncertainties.}
\label{tab:babar-sll}
\begin{tabular}{ccc}
Channel & This measurement & SM theory \\ \hline
\multicolumn{3}{c} {$1 < q^2 < 6$ GeV$^2$} \\ \hline
 $B\to X_s \mu^+\mu^-$ & \result{0.66}{0.82}{0.76}{0.30}{0.24}$\pm 0.07$ & $1.59 \pm 0.11 $ \\
 $B\to X_s e^+e^-$     & \result{1.93}{0.47}{0.45}{0.21}{0.16}$\pm 0.18$ & $1.64 \pm 0.11 $ \\
 $B\to X_s \ell^+\ell^-$ & \result{1.60}{0.41}{0.39}{0.17}{0.13}$\pm 0.18$ &  \\ \hline
\multicolumn{3}{c} {$14.2 < q^2$ GeV$^2$} \\ \hline
 $B\to X_s \mu^+\mu^-$ & \result{0.60}{0.31}{0.29}{0.05}{0.04}$\pm 0.00$ & $0.25^{+0.07}_{-0.06} $ \\
 $B\to X_s e^+e^-$     & \result{0.56}{0.19}{0.18}{0.03}{0.03}$\pm 0.00$ &  \\
 $B\to X_s \ell^+\ell^-$ & \result{0.57}{0.16}{0.15}{0.03}{0.02}$\pm 0.0$ &  \\
\hline
\end{tabular}
\end{center}
\end{table}
In the experimental measurements, the first uncertainty listed is statistical, the second is experimental systematic and
the third is model-dependent systematic related to the extrapolation to the full hadronic mass spectrum and inclusion
of the missing modes. These results are compatible with expectations, although the BF in the high-\qsq\ region is
approximately $2\sigma$ above the SM value. We note that it also is $2\sigma$ from the most favoured value of the
beyond-SM contribution $C_9^{BSM}$, which has been proposed to explain recent observations in the channel $B^0\to
K^{*}\mu^+\mu^-$ by the LHCb Collaboration~\cite{ref:lhcb-kll}. 

The CP asymmetry is also determined from the event yields, using the 14 self-tagging modes {\em i.e.}, excluding the
modes with $X_s \in \{K_S^0, K_S^0\pi^0,K_S^0\pi^+\pi^-\}$. No model-dependent extrapolation of signal rates is
attempted, so the \acp\ result pertains only to the modes utilized. The result obtained
\begin{equation}
\acp \equiv \frac{\Gamma_{\bar B} - \Gamma_B}{\Gamma_{\bar B} + \Gamma_B} = 0.04 \pm
0.11_\mathrm{stat}\pm0.01_\mathrm{syst}
\end{equation}
is consistent with the SM prediction of very small CP asymmetry~\cite{ref:sll-acp-theory}.

\subsection{Forward-backward lepton asymmetry in \bxsll}

Belle has made the first measurement of the forward-backward lepton asymmetry in inclusive
\bxsll\ decays~\cite{ref:belle-sll}. The asymmetry, defined as follows:
\begin{equation}
A_\mathrm{FB}(q^2_\mathrm{min},q^2_\mathrm{max}) = \frac{\int^{q^2_\mathrm{max}}_{q^2_\mathrm{min}} dq^2 \int_{-1}^1
      d\cos\theta \,\mathrm{sgn}(\cos \theta)\frac{d^2\Gamma}{dq^2 d\cos\theta} }
{\int^{q^2_\mathrm{max}}_{q^2_\mathrm{min}} dq^2 \int_{-1}^1 d\cos\theta \frac{d^2\Gamma}{dq^2 d\cos\theta} },
\end{equation}
is sensitive to NP effects. Here $\theta$ is the angle between the
positive (negative) lepton and the B meson momentum in the $\ell^+\ell^-$ center-of-
mass frame in $\bar B^0$ or $B^-$ ($B^0$ or $B^+$) decays. 

\begin{figure}[htb]
\centering
\includegraphics[width=.8\textwidth]{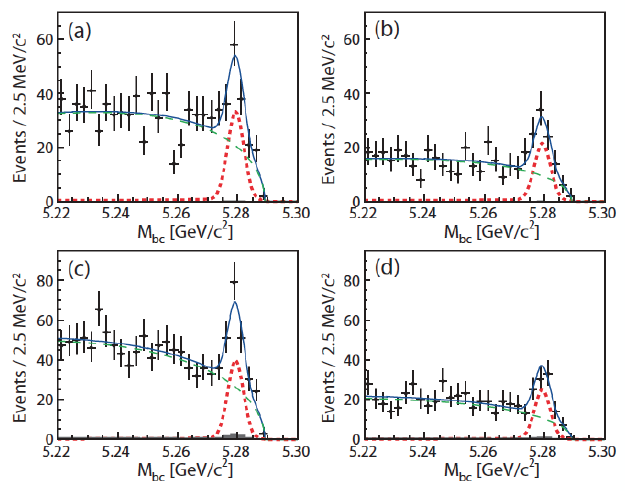}
\caption{The B candidate mass $M_\mathrm{bc}$ for inclusive \bxsll\ events. The top row shows the \epem\ channels, while
the bottom row displays the \mpmm\ modes. The forward-going sample is shown on the left, while the backward-going events
are on the right.}
\label{fig:afb-yields}
\end{figure}

The Belle analysis, based on their dataset of 772 million \BB\ pairs, employs the sum-of-exclusives technique, using 10
self-tagging modes, which account for roughly 50\% of the total rate. The event selection is standard: PID for charged
particles, multivariate classifier (neural net, in this case) to suppress continuum backgrounds and explicit charmonium
vetoes. The quantity \afb\ is measured in 4 \qsq\ bins. Events are divided into forward and backward sub-samples and the
signal yields are extraced by a fit to the B candidate mass $M_\mathrm{bc}$ (Fig.~\ref{fig:afb-yields}). 

\begin{figure}[htb]
\centering
\includegraphics[height=2.5in]{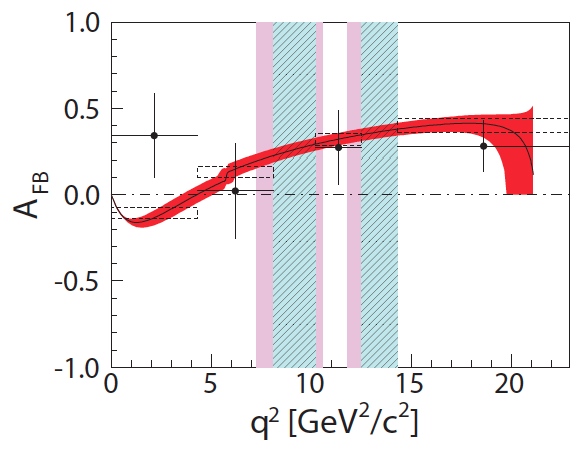}
\caption{Forward-backward lepton asymmetry in inclusive \bxsll\ events. The black points represent the measurement,
  while the red band is the theoretical prediciton. The charmonium vetoes are also shown.}
\label{fig:sll-afb}
\end{figure}

The raw yields are
corrected for efficiency, which varies considerably over \qsq\ and $\cos\theta$. Simulated events are used to derive the
correction factors. Figure~\ref{fig:sll-afb} shows the \afb\ results as a function of \qsq. The results are consistent with
the SM expectation~\cite{ref:afb-theory}. Table~\ref{tab:afb} reports the results in numerical form.
This is the first measurement of \afb\ for inclusive \bxsll.
\begin{table}[htb]
\caption{\label{tab:afb}Fit results for the four $q^2$ bins.
For $A_{\rm FB}$,
the first uncertainty is statistical
and the second uncertainty is systematic.
SM predictions for the ${A}_{\rm FB}$ values are 
from~\cite{ref:afb-theory}. The units of \qsq\ are GeV$^2$.
}
{\footnotesize
\begin{tabular}{cc|cccc}
                                      &           & 1st bin                    & 2nd bin                    & 3rd bin                    & 4th bin                      \\ \hline
\multirow{2}{*}{$q^2$ range} & ($B \rightarrow X_s e^+ e^-$) & \multirow{2}{*}{[0.2,4.3]} & [4.3,7.3]                  & [10.5,11.8]                & \multirow{2}{*}{[14.3, 25.0]} \\
& ($B \rightarrow X_s \mu^+ \mu^-$)   &                            & [4.3,8.1]                  & [10.2,12.5]                &                                    \\ \hline
\multicolumn{2}{c|}{${A}_{\rm FB}$}          & $0.34  \pm 0.24 \pm 0.02$  & $0.04 \pm 0.31 \pm 0.05$   & $0.28 \pm 0.21 \pm 0.01$ & $0.28 \pm 0.15 \pm 0.01$ \\
\multicolumn{2}{c|}{${A}_{\rm FB}$ (theory)} & $-0.11 \pm 0.03$           & $0.13 \pm 0.03$            & $0.32 \pm 0.04$          & $0.40 \pm 0.04$          \\
\end{tabular}
}
\end{table}

\section{Conclusions}

Inclusive measurements of the decays \bxsg\ and \bxsll\ are important tools for constraining models of New
Physics. The measurements presented herein use the full datasets from the Babar and Belle experiments and represent the
state of the art regarding these channels at the B-factories. While inclusive \bxsg\ has been well-studied at the B factories, measurements of the \bxsll\ channel are still
in their infancy. Because of the difficulties of making inclusive measurements at hadron colliders, further progress
will have to await the advent of the Belle II experiment, which is expected to come online within a few years. The
prospects at a very high-luminosity \epem\ machine are very exciting: the large statistics will lead to precision
measurements in \bxsll, while full reconstruction of tag-side hadronic B decays will lead to new possibilities in
\bxsg\ and the Cabbibo-suppressed decay \bxdg. 

\Acknowledgements
I am grateful to the organizers of FPCP 2014 in Marseille. They did a wonderful job of organizing a stimulating and
interesting conference.

\end{document}